\documentclass[aps,prb,superscriptaddress,floatfix]{revtex4}
\usepackage{graphicx,bm}
\usepackage{dcolumn}
\usepackage{bm}
\usepackage{latexsym}
\usepackage{amssymb}
\usepackage{amsfonts}
\usepackage{amsmath}
\usepackage{fancyhdr}
\usepackage{epsfig,epsf}
\newcommand{\vDe}{\vec \Delta}
\newcommand{\vm}{\vec{m}}

\usepackage{color}

\definecolor{MyDarkBlue}{rgb}{0,0.08,0.45}

\def\bx{{\bf x}}

\begin{document}
\title{Current-Induced Torques in Magnetic Metals: Beyond Spin Transfer}
\author{P. M. Haney}
\email{haney411@physics.utexas.edu}
\homepage{http://www.ph.utexas.edu/~haney411/paulh.html}
\affiliation{ Department of Physics, The University of Texas at Austin, Austin,
Texas, 78712-0264, U.S.A. }

\author{R. A. Duine}
\email{duine@phys.uu.nl}
\homepage{http:www.phys.uu.nl/~duine}
\affiliation{Institute for Theoretical Physics, Utrecht
University, Leuvenlaan 4, 3584 CE Utrecht, The Netherlands}

\author{A. S. N\'{u}\~{n}ez}
\email{alvaro.s.nunez@gmail.com} \affiliation{Departamento de
F\'isica, Universidad de Santiago de Chile, Casilla 307, Santiago,
Chile. }

\author{A. H. MacDonald}
\email{macd@physics.utexas.edu}
\homepage{http://www.ph.utexas.edu/~macdgrp}
\affiliation{ Department of Physics, The University of Texas at Austin, Austin,
Texas, 78712-0264, U.S.A. }

\date{\today}


\begin{abstract}
Current-induced torques on ferromagnetic nanoparticles and on
domain walls in ferromagnetic nanowires are normally understood in
terms of transfer of conserved spin angular momentum between
spin-polarized currents and the magnetic condensate.  In a series
of recent articles we have discussed a microscopic picture of
current-induced torques in which they are viewed as following from
exchange fields produced by the misaligned spins of current
carrying quasiparticles.  This picture has the advantage that it
can be applied to systems in which spin is not approximately
conserved.  More importantly, this point of view makes it clear
that current-induced torques can also act on the order parameter
of an antiferromagnetic metal, even though this quantity is not
related to total spin.  In this informal and intentionally
provocative review we explain this picture and discuss its
application to antiferromagnets.
\end{abstract}

\maketitle

\section{Spin Transfer Torques}
The study of spin-transfer torques began in 1996 when John
Slonczewski \cite{slonczewski} and Luc Berger \cite{berger}
independently predicted that magnetization dynamics can be induced
by current in circuits containing noncollinear magnetic elements.
Berger's paper focused on the emission of spin waves as the source
of magnetic excitations, while Slonczewski invoked conservation of
spin angular momentum to infer magnetic dynamics. Slonczewski's
observation that a net spin-current flux into a volume of magnetic
material implies that a torque acts on the magnetization in that
volume is the key idea for most theories of current-induced
torques (CIT).  Over the past decade many experiments have
confirmed \cite{tsoi1,tsoi2,katine,albert,pufall,kiselev,grollier}
Slonczewski's predictions.  There has also been theoretical
progress, elaborating on Slonczewski's ideas and developing
techniques which shed light on their implications for particular
materials combinations and geometries.  For example, Stiles and
Zangwill explicitly exhibited all of the sources of net
spin-current flux, namely spin-dependent transmission, spin
precession, and rotation of reflected and transmitted
spins.\cite{stilesZangwill} Brataas {\em et al.} have formulated a
general theory in which spin-dependent interface conductances,
calculated from first principles or extracted from experimental
data, can be combined using a generalized set of Kirchhoff laws to
predict transport properties and magnetization dynamics in a
circuit containing noncollinear magnetic elements.\cite{brataas}
Other approaches include solving generalized Boltzmann\cite{xiao}
or spin diffusion equations.\cite{zhang}  All of this theoretical
work is directed towards evaluation of the net spin-current into a
volume of material. Since the underlying systems possess
conservation of total spin angular momentum, any theory for these
systems must respect this global conservation law. For systems in
which total spin is conserved, the relationship between net spin
current and torques is very general.  Appealing to spin
conservation enables reliable predictions to be made about
current-induced magnetization dynamics without having to specify
which spins form the macroscopic magnetization or the microscopic
origin of the effective magnetic fields which cause them to
precess.  The conservation laws assure that if one does the
``bookkeeping" of spin properly, the current-induced torque acting
on a volume may be inferred. We therefore sometimes refer to the
point of view which utilizes powerful conservation-law
consequences as the {\em bookkeeping theory} of spin-transfer.
This point of view has had qualitative and quantitative success in
describing experiments on spin-transfer in spin valves systems
which are composed of several ferromagnetic nanoparticles, and on
spin-transfer induced domain wall motion in ferromagnetic
nanowires.

The bookkeeping theory of spin-transfer torques raises two
questions which often need not be answered explicitly and which
initially drew our interest to the issue of current-induced
torques.  In attempting to provide answers to these questions that
we find satisfying, we have been led to the theoretical picture of
current-induced torques described below.

\noindent {\em i. What is the distinction in the spin-transfer
picture between the electrons that carry current and the electron
spins that compose the magnetic condensate?}  This question is
particularly relevant for the transition metal systems which are
the workhorses of metal spintronics since we know that both
$s$-like and $d$-like orbitals must be treated as itinerant. There
is no clean distinction between the electrons which contribute the
moments that order and the electrons that carry current.

\noindent {\em ii. Does spin-transfer occur in systems with strong
spin-orbit coupling?}  Because of spin-orbit coupling, spin
angular momentum is never really conserved, even in perfect
crystals that have no disorder.  How strong does spin-orbit
coupling need to be to weaken or eliminate the spin-transfer
effect?  This question is particularly relevant to ferromagnetic
semiconductors like (Ga,Mn)As in which the spin-orbit coupling
strength is\cite{GaMnAsRMP} comparable to the magnetic exchange
energy and to the Fermi energy.

\section{Current-induced torques and non-equilibrium spin-densities}

In our picture spin-transfer torques arise as follows:  The
spin-density of electrons near the Fermi energy of a magnetic
metal is altered when they carry a current through a non-collinear
magnetic environment.  The change occurs as they realign their
spins to sample the exchange field of the ferromagnet and thereby
steer their spin orientations to match their non-collinear
environment.  This change in spin-density leads to a change in the
exchange field of the ferromagnet, which causes spins in orbitals
far from the Fermi level to precess.  We will refer to this as the
{\em current-induced torque} picture of spin-transfer.  Although
consistent with the bookkeeping theory of spin-transfer for
systems in which total spin is conserved, it suggests that the
phenomena is more general. One implication as we discuss below is
that current-induced torques drive order parameter dynamics in
antiferromagnetic metals.

\subsection{Current-induced torques}
We first introduce some notation and provide a general
orientation. The formalism we describe is applicable to any mean
field theory, but in this article we consider primarily systems
described by the local spin-density approximation (LSDA) of
density functional theory (DFT). The notation we use below is
appropriate for a Hamiltonian in a tight-binding or local orbital
basis.  We separate both the single-particle Hamiltonian and the
density matrix into spin-dependent and spin-independent
contributions:
\begin{eqnarray}
\label{eq:spindependence}
\rho_{i'i} &=& \frac{1}{2} ~\big[ \rho^{(0)}_{i'i}  + \vec{m}_{i'i} \cdot {\vec \tau} \big], \nonumber \\
{\rm H}_{i'i} &=& {\rm H}^{(0)}_{i'i}  - \frac{1}{2}
\vec{\Delta}_{i'i} \cdot {\vec \tau}.
\end{eqnarray}
where $\vec \tau$ is the vector of Pauli spin matrices and $i',i$
are site or orbital indices.  In DFT, ${\rm H}_{i'i}$ is the
Kohn-Sham single-particle Hamiltonian.  The notation for the
spin-dependent part of the ${\rm H}$ is chosen to emphasize that
it produces a spin-splitting $\Delta$ when it is orbital
independent, as is often assumed in simple toy models of a
ferromagnetic metal. (Note that $\vec{m}$ and $\vec{\Delta}$ are
in general complex for orbital off-diagonal elements, {\em i.e.}
for $i \neq i'$; any Hermitian matrix may be uniquely written in
the above form.) In LSDA, the interaction contribution to ${\vec
\Delta}$ and ${\vec m}$ are related locally at each point in space
according to
\begin{equation}
\label{eq:lsda} {\vec \Delta}(\vec r)=\Delta_0(n(\vec r),m(\vec
r)) \; {\hat{m}}.
\end{equation}
 where $n$ and ${\vec m}= m \, \hat{m}$ are the
local charge and spin-densities, respectively, and $\Delta_0$ is
some parameterization of the exchange-correlation potential. Note
that ${\vec \Delta}(\vec r)$ acts like an effective-magnetic field
experienced by the Kohn-Sham quasiparticles. (In Eq.
(\ref{eq:lsda}), $\vec \Delta(\vec r)$ and $m(\vec r)$ are
functions in real space, while in Eq. (\ref{eq:spindependence}),
$\Delta_{i,i'}$ represents the $i,i'$ matrix element of the real
space potential $\Delta(\vec r)$, and $\vec m_{i,i'}$ the
spin-dependent part of density matrix in orbital space.)  A local
relationship in space does not imply proportionality of the
orbital representation matrix elements.  In particular there are
strong exchange interactions between $s$-like and $d$-like
orbitals.

We define the difference between spin-densities in the presence
and in the absence of a current as the non-equilibrium
spin-density, denoted by ${\vec m}^{tr}$, and the corresponding
difference in exchange-correlation potentials is denoted by ${\vec
\Delta}^{tr} = \Delta_0(n,m)~ {\vec m}^{tr}/|m|$.  This expression
assumes that $|{\vec m}^{tr}|/m$ is small, something that is valid
for any reasonable current strength.  In a circuit with a
noncollinear magnetic configuration, the contribution to the local
exchange-correlation effective magnetic field from the
non-equilibrium quasiparticles will in general point in a
different direction than the magnetic condensate, and this
misalignment is responsible for the ensuing torque on the magnetic
condensate - the spin-transfer torque.\cite{nunez}  The
non-equilibrium spin-density driven by a source to drain bias
voltage in a specific nano-scale circuit can be evaluated
theoretically using non-equilibrium Greens function techniques, as
we describe below. For a bulk magnetic metal with a smooth
spin-texture, the non-equilibrium spin-density can be evaluated by
treating both the spatial variation of magnetization direction and
the electric field which drives a bulk current perturbatively.

\begin{figure}[h]
\begin{center}
\vskip 0.25 cm
\includegraphics[width=3.5in,angle=0]{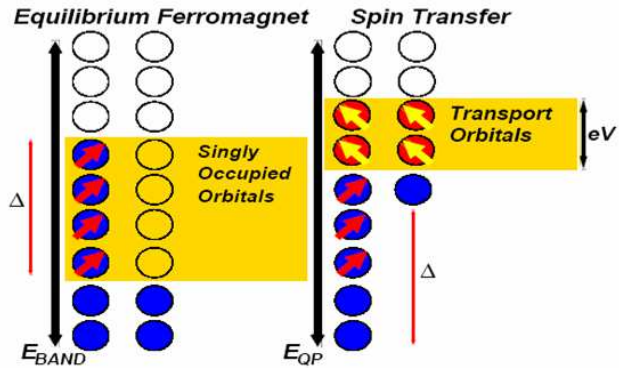}
\vskip 0.25 cm \caption{Left panel: Ground state of a metallic
ferromagnet. The low-energy collective degree of freedom, the
magnetization direction, is the spin orientation of singly
occupied orbitals. Right panel: Quasiparticles experience a strong
exchange field $\vec \Delta$ that brings majority and minority
spins into equilibrium. Because this field is parallel to the
magnetization it does not produce a torque. In an inhomogeneous
ferromagnet, the spin orientation of the transport orbitals (in a
window of width eV at the Fermi energy) must differ from the
direction of the exchange-correlation potential $\Delta$. The
current-transfer torque is then produced by the transport-orbital
contribution to the exchange field.} \label{fig:lsda}
\end{center}
\end{figure}

\begin{figure}[h]
\begin{center}
\vskip 0.25 cm
\includegraphics[width=3.5in,angle=0]{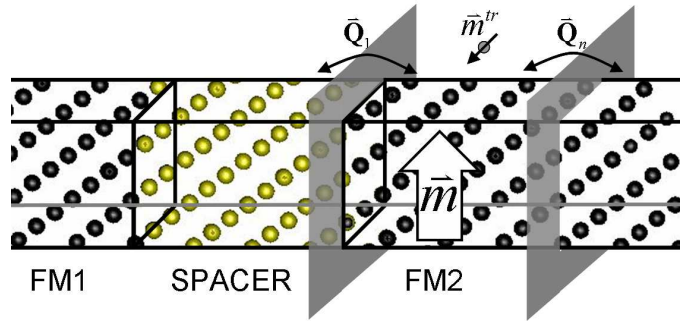}
\vskip 0.25 cm \caption{Illustration of the relation between
spin-current fluxes ${\bf \vec Q}$ and non-equilibrium
spin-densities $\vm^{tr}$. The planes through which spin-currents
${\bf \vec Q}$ are evaluated are between layers labelled 0 and 1,
and $n-1$ and $n$. } \label{fig:spinFlux}
\end{center}
\end{figure}

As an example, consider the spin valve structure of two
ferromagnets separated by a spacer. If a bias is applied, there is
a component of the non-equilibrium spin-density which is
perpendicular to the plane spanned by the two ferromagnets.
Because this spin-density is not parallel to the
exchange-correlation potential $\Delta$, the transport orbital
spins precess as they move through the circuit to accommodate the
change in exchange-field orientation. The magnetization of both
layers will precess around the local out-of-plane exchange field
generated by this non-equilibrium spin-density, and this
precession is the one which leads to current-induced magnetization
switching (CIMS) (see Fig. (\ref{fig:spinFlux})).

In the case where total spin is conserved, this view is
essentially equivalent to the bookkeeping theory of spin-transfer,
as can be seen by evaluating $d\vm^{tr}/dt$ for a current-carrying
quasiparticle with source to drain scattering boundary conditions:

\begin{eqnarray}
\frac{d \vm^{tr}}{dt} = \frac{1}{i\hbar}\left[\vm^{tr},H\right] =
\left({\vec Q}_1 - {\vec Q}_n\right) + \frac{1}{\hbar}{\rm
Tr}\left[{\rm Re}\left[\vm^{tr} \times {\vDe}\right]\right]~.
\label{eq:dmdt}
\end{eqnarray}
The above relation can be obtained by evaluating the commutator
directly and using the commutation properties of Pauli spin
matrices. From Eq. (\ref{eq:spindependence}), $\vm^{tr}$ is the
spin-dependent part of the scattering state density matrix and
$\vDe$ is the spin-dependent part of the Hamiltonian, each of
which is expanded in the 3 Cartesian components of Pauli matrices
$\tau_x,\tau_y,\tau_z$.  The trace in the second term is over
orbitals in the subsystem (in the example shown by Fig.
(\ref{fig:spinFlux}), the subsystem consists of all orbitals
between planes 1 and $n$), and the spin-dependent part of matrices
are multiplied in cross product form.

The spin-current operator ${\vec Q}_{i}$ is defined as the
spin-current that flows between sites $i$ and $i+1$, and is given
by
\begin{eqnarray}
{\vec Q}_{i} = \frac{1}{\hbar}\mathop{\sum_{k \leq i}}_{j
> i} {\rm Tr}\left[{\vec\tau}\left(\rho_{kj}H_{jk} - H_{kj}\rho_{jk}\right)\right] ~. \label{Q}
\end{eqnarray}

\noindent In Eq. (\ref{Q}) the trace is over spin space, and
$H_{ij}$ ($\rho_{ij}$) is the $2 \times 2$ spin matrix of the
$i,j$-orbital component of $H$ ($\rho$). The second term of Eq.
(\ref{eq:dmdt}) represents the torque present on the quasiparticle
due to its misalignment with the magnetic condensate.  In steady
state transport, the left hand side of Eq. (\ref{eq:dmdt})
vanishes, and so the divergence of the spin-current - or the net
spin-current flux, is equal to the quasiparticle-condensate
current-induced torque. The equation of motion for the magnetic
condensate is \cite{antropov}:
\begin{eqnarray}
\frac{d \vm}{dt} = - \vm  \times \vDe = - \vm^{tr}\times\vDe ~.
\label{eq:dmdt2}
\end{eqnarray}
From Eq. (\ref{eq:dmdt}), this implies that the torque on $\vm$
can be calculated with either $\vm^{tr}$ (the non-equilibrium
spin-density) or $({\vec Q}_1 - {\vec Q}_n )$ (the net
spin-current flux), verifying the consistency between the
approaches. In identifying the current-induced torques as arising
from non-equilibrium spin-densities, we have accomplished more
than simply rephrasing the bookkeeping argument. We have
identified the underlying microscopic interaction that is
responsible for current-induced torques.  The same mechanism is
operational in systems in which spin is not conserved. In
addition, this microscopic view allows for the evaluation of local
torques on individual atoms which can also drive the order
parameter of antiferromagnets.  The current-induced torque which
acts on the order parameter of a volume of antiferromagnetic
material is not related to the net spin-current flux into that
volume. We comment more on these systems in Sec.
\ref{sec:ferrimagnets}.

\subsection{Current induced torques and exchange interactions.}
So far we have identified the current-induced torque as resulting
from the misalignment of the magnetic condensate with the exchange
field contribution of non-equilibrium quasiparticles near the
Fermi energy.  This misalignment may seem out of place in view of
Eq. (\ref{eq:lsda}) - in the ground state, the total spin and
exchange field are aligned.  However, even in equilibrium systems
(systems that carry no charge current) with magnetic excitations
(such as spin waves), the magnetic dynamics can be determined from
an expression similar to Eq. (\ref{eq:dmdt2}) (assuming the
excitation energy is ``small" compared to other characteristic
energies). In this case the Hamiltonian $H' =
{H^{(0)}}'-1/2\left(\vec{\Delta}'\cdot\vec\tau\right)$ describing
the magnetic excitation is constructed ``by hand" starting from
the ground state Hamiltonian (by, for example, applying a
space-dependent spin rotation operator which describes a spin wave
imposed on a collinear ground state Hamiltonian). The resulting
{\em non self-consistent} density matrix
$\rho'=1/2\left({\rho^{(0)}}'+\vec{m}'\cdot\vec\tau\right)$ can
then easily be calculated.  The torque on the magnetic system is
then given by $ {\rm Tr} \left[{\rm Re}\left[\vm'  \times
\vDe'\right]\right]$, as in Eq. (\ref{eq:dmdt2}), and yields
proper values for magnetic exchange energies and spin wave
dispersion relations.  Indeed, it can be shown that this approach
to calculating properties of magnetic excitations is equivalent to
previous approaches\cite{leichtenstein} which calculate $\delta
E/\delta \hat m$ - the change in energy associated with magnetic
excitations $\delta \hat m$.\cite{paulthesis}  (As a technical
note, we remark that the evaluation of $\delta E/\delta \hat m$,
as well as our evaluation of torques, relies on an adiabatic
condition for magnetic dynamics.  The adiabatic approximation
follows from the fact that electronic times scales are much faster
than the time scales which characterize collective magnetic
dynamics.\cite{antropov}) Having established the close
relationship between torques and variations in energy for
excitations of equilibrium systems, one is naturally led to ask if
such a relation exists for non-equilibrium systems.  We address
that point in Sec. \ref{sec:energy}.

\subsection{How to calculate non-equilibrium spin-densities}

We now briefly describe the technique we use to evaluate
spin-densities and provide some references. This in some sense
represents a technical point outside the focus of this article
(although certainly a point of practical importance). We adopt a
Landauer-Buttiker approach for describing non-equilibrium
transport.  In this approach, the bias voltage is represented by
placing the system in contact with particle reservoirs with
chemical potentials $\mu_S = \epsilon_F + e V_B/2$ and $\mu_D =
\epsilon_F - e V_B/2$ in source and drain respectively.  When the
system Hamiltonian and its coupling to source and drain electrodes
is time-independent (an assumption which follows from the
adiabatic approximation described above), electrons with energies
inside of the transport window $\mu_D < E < \mu_S$ solve a
time-independent Schrodinger equation with incident-from-source
scattering boundary conditions.  In the Landauer-Buttiker
approach, the conductance is simply proportional to the
quantum-mechanical transmission probability of the scattering
state.  There are a number of techniques available to solve this
system. We employ a non-equilibrium Green's function (NEGF) method
\cite{datta} combined with density functional theory.\cite{taylor}
The NEGF formalism yields all quantities of interest: the
conductance, and the contribution to the density matrix from the
non-equilibrium scattering states, from which $\vm^{tr}$ is
determined.  A useful introduction to the NEGF formalism can be
found in Ref. ~\onlinecite{dattapaper}, and a more formal account
is given in Ref. ~\onlinecite{haug} . More details of how NEGF can
be used to find current-induced torques can be found in Ref.
~\onlinecite{haneyFM}.

\section{Reactive and dissipative torques}
\label{sec:energy}

In the previous section we described how current-induced torques
arise as a consequence of the interaction between the magnetic
condensate and the spin-density of non-equilibrium quasiparticles,
a point of view that is an extension of the notion of spin-torques
arising from spin-currents.  The exchange-correlation field of a
non self-consistent spin-density may also be associated with the
torques arising from magnetic stiffness, and in this context these
torques are related to the variation of an energy functional.  We
now comment on the possibility of finding current-induced torques
from an energy functional, which is of particular relevance in
view of the ongoing discussion regarding the form of magnetization
damping.\cite{stilesdamping,smith}

The Landau-Lifshitz-Gilbert equation describes the dynamics of the
magnetization.  When spin-transfer torques are included, it is:

\begin{equation}
\label{eq:LLG}
  \frac{\partial {\hat m}}{\partial t} = {\hat m} \times
  \left( -\frac{\delta E[{\hat m}]}{\hbar \delta {\hat m}}
  \right) - \alpha {\hat m} \times \frac{\partial {\hat m}}{\partial
  t} + \left. \frac{\partial {\hat m}}{\partial t} \right|_{\rm
  STT}~.
\end{equation}
Here ${\hat m}$ is the magnetization vector normalized to unit
magnitude. The first term on the right hand side describes
precession of the magnetization around an effective field, written
here as the functional derivative \footnote{The functional
derivative $\delta E[\hat m]/(\delta \hat m)$ is defined by the
energy variation $\delta E = \int d\bx [\delta E/(\delta \hat m
(\bx))] \cdot \delta \hat m (\bx)$ of the energy functional.
Writing the energy in terms of an energy density ${\mathcal E}
(\hat m)$, {\em i.e.}, $E[\hat m] = \int d\bx {\mathcal E} (\hat m
(\bx))$, we have that \[
 \frac{ \delta E[\hat m]}{\delta \hat m} = \frac{\partial {\mathcal E}}{\partial \hat m}
  - \frac{\partial}{\partial x_\gamma} \frac{\partial {\mathcal E}}{\partial (\partial \hat m/\partial
  x_\gamma)}~,
\] where a sum over the repeated index $\gamma \in \{x,y,z\}$ is again implied.  See also the introductory article for a discussion of functional derivatives.}
of the energy. In the so-called micromagnetic theory
used to describe smooth spin-textures, the energy functional
contains contributions from Zeeman coupling of the magnetization
to an external magnetic field, anisotropy energy contributions due
to spin-orbit coupling and magnetostatic interactions which
violate spin conservation, as well as the energy cost of
magnetization variation which is referred to in the context of
micromagnetic theory as the exchange energy. (As experts are well
aware, the theory of magnetism refers to many different things as
exchange energies.  The many uses of this word is fitting, since
magnetism is always intimately related to electronic exchange
processes, but it can be confusing.) In Eq.(~\ref{eq:LLG}), the
term proportional to $\alpha$ is the Gilbert damping term.

For smooth spin-textures, the spin-transfer torque terms can be
expressed in terms of the leading order of a gradient (spatial
derivative) expansion:
\begin{equation}
\label{eq:sttcont}
  \left. \frac{\partial {\hat m}}{\partial t} \right|_{\rm
  STT} = -({\bf v}_s \cdot \nabla) {\hat m} + \beta {\hat m}
  \times
  ({\bf v}_s \cdot \nabla) {\hat m}~,
\end{equation}
where the velocity ${\bf v}_s=-a^3P{\bf j}/|e|$ is proportional to
the electric current density ${\bf j}$.  Here $a$ is introduced as
the lattice constant of a fictional lattice of unit magnetic
moments representing a magnetization density $a^{-3}$, and $P$ is
the spin polarization fraction of the current. Finally, the
electron charge is denoted by $-|e|$. The first term in
Eq.~(\ref{eq:sttcont}) is known as the adiabatic spin-torque. When
${\bf v}_s$ is defined as above, this term is just the gradient
expansion limit of the spin-conserving spin-transfer torque
discussed in the previous section. The second term proportional to
the dimensionless parameter $\beta$ is commonly (and perhaps
inappropriately - see below) referred to as the nonadiabatic
spin-torque.  Since $\hat{m}$ is a unit vector, it is
perpendicular to its time derivative $\dot{\hat{m}}$. It follows
that when both $P$ and $\beta$ are treated as phenomenological
parameters, Eq. (\ref{eq:sttcont}) is quite general, assuming only
that the current-induced torque is linear in current.  From a
symmetry point of view, the first spatial derivative spin-transfer
torque terms in the generalized Landau-Lifshitz equations are
allowed because current-flow breaks inversion symmetry.  The
micromagnetic {\em exchange} term, which is proportional to the
second spatial derivative of the magnetization, is the leading
order term in a gradient expansion in the absence of current. Eq.
(\ref{eq:sttcont}) can be derived
microscopically,\cite{rossier,duine,kohno,tserkovnyak} including
the spin-transfer torque terms, by using non-linear response
theory to describe the response of magnetization to external
fields in the presence of a transport current.

The analogous equation for the free nanomagnet in a spin-valve system is
\begin{equation}
\label{eq:sttslonc}
  \left. \frac{\partial {\hat m}}{\partial t} \right|_{\rm
  STT} = -g j {\hat m} \times \left( {\hat m}_p \times {\hat m} \right)
  - \beta g j \left( {\hat m}_p \times {\hat m} \right)~,
\end{equation}
with ${\hat m}_p$ the magnetization direction of the pinned
magnet. In this equation $g \sim a^3 P/(2|e|\ell)$ with $\ell$ the
length of the free ferromagnet in the direction of current flow.
In this section we follow one common usage by referring to the
first term as the Slonczewski spin-torque, and to the second term
as the effective-field spin-torque; physically $g$ represents the
efficiency of the Slonczewski spin-torque.  An important
observation is that Eqs.~(\ref{eq:sttcont}) and
(\ref{eq:sttslonc}) are two sides of the same coin. If we put
${\hat m}_p ={\hat m} (r-dr)$ in Eq.~(\ref{eq:sttslonc}), with $r$
the coordinate in the direction of the current, we obtain, after
expanding to lowest order in $dr$ the result in
Eq.~(\ref{eq:sttcont}).  In the case of spin-valve structures we
know from microscopic theory that the approximate expression for
$g$ has corrections from spin-dependent reflection and other
effects that are easily captured by microscopic interface
calculations.\cite{stilesZangwill,brataas}

A useful classification of the terms in Eq.~(\ref{eq:LLG}) follows
from an examination of how they behave under time reversal
operations.  Letting $t \mapsto -t$, we have that ${\hat m}, {\hat
m}_p \mapsto -{\hat m}, -{\hat m}_p$, and ${\bf v}_s, j \mapsto -
{\bf v}_s, -j$. Furthermore, $E[-{\hat m}] = E[{\hat m}]$. After
implementing this operation in Eq.~(\ref{eq:LLG}) we observe that
there are two kinds of terms, reactive terms that are even under
time reversal and dissipative terms that are odd. (Readers should
be aware that there exist different points of view regarding this
issue, found in Ref. ~\onlinecite{saslow} and
~\onlinecite{smith}.) The terms reactive and dissipative originate
from linear response theory in which they play a similar role,
distinguishing response that is in phase with a periodic driving
force from response that is out of phase and therefore
dissipative.  The first and second terms on the right hand side of
Eqs. (\ref{eq:sttcont}) and  (\ref{eq:sttslonc}) respectively
represent corrections to the reactive and dissipative terms in Eq.
(\ref{eq:LLG}) due to current flow. In the smooth texture
(continuum) limit this is obvious from a microscopic point of
view, because the terms proportional to $\alpha$ and $\beta$
emerge from microscopic linear-response theory as dissipative
parts of the spin-density spin-density response function in the
presence of current.\cite{duine} To many readers the
classification of the Slonczewski spin-torque as reactive may come
as a surprise: after all, the Slonczewski spin-torque competes
with the Gilbert damping in current-driven magnetization reversal.
The latter phenomenon is dynamic, however, and does not imply that
the Slonczewski spin-torque is dissipative.  This classification
is consistent with our picture that current-induced torque
phenomena primarily reflect precession around a transport-induced
contribution to the exchange-correlation effective magnetic field.

The fact that both the adiabatic spin-torque and the Slonczewski
spin-torque are both reactive triggers the question whether they
can be derived from an energy functional. Indeed, as we show
below, the action
\begin{eqnarray}
\label{eq:actionwithspintorques}
 {\mathcal  A} [{\hat m}] = \int d t \left\{ -\left[ \int \frac{d\bx}{a^3}
 \hbar v_{s,\gamma}  A_{\gamma'}({\hat m} (\bx,t)) \nabla_\gamma m_{\gamma'} (\bx,t)  + \hbar {\bf A} ({\hat m} (\bx,t)) \cdot \frac{\partial {\hat m} (\bx,t)}{\partial t}\right]
  - E [{\hat m}] \right\}~,
\end{eqnarray}
with ${\bf A}$ the vector potential of a magnetic monopole
determined by
\begin{equation}
\label{eq:vecpotmonopole}
  \epsilon_{\gamma\gamma'\gamma''} \frac{\partial A_{\gamma''}}{\partial \hat
  m_{\gamma'}}= \hat m_\gamma~.
\end{equation}
that enforces spin quantization,\cite{auerbach} reproduces, upon
variation, the equation of motion in Eq.~(\ref{eq:sttcont}) for
$\alpha =\beta=0$.\cite{bazaliy} We note that in the above
equation $\epsilon_{\gamma\gamma'\gamma''}$ is the antisymmetric
Levi-Civita tensor and sums over repeated Cartesian indices
$\gamma,\gamma',\gamma'' \in \{x,y,z\}$ are implied.

To obtain the equation of motion corresponding to the action in
Eq.~(\ref{eq:actionwithspintorques}) we have to calculate the
variation of the action, {\em i.e.}, $\delta {\mathcal A}/\delta
\hat m_\gamma $ and put it equal to zero. We find in first
instance that
\begin{equation}
\label{eq:nearlythere}
  0=\frac{\delta {\mathcal A}}{\delta \hat
m_\gamma}= \hbar \left( \frac{\partial A_\gamma}{\partial \hat
m_{\gamma'} }-\frac{\partial A_{\gamma'}}{\partial \hat m_{\gamma}
}\right) \left(\frac{\partial}{\partial t} + v_{s,\gamma''}
\frac{\partial }{\partial x_{\gamma''}} \right) \hat m_{\gamma'} -
\frac{\delta E}{\delta \hat m_\gamma}~.
\end{equation}
Using Eq.~(\ref{eq:vecpotmonopole}) we have that
\begin{equation}
  \frac{\partial A_\gamma}{\partial \hat
m_{\gamma'} }-\frac{\partial A_{\gamma'}}{\partial \hat m_{\gamma}
} = -\epsilon_{\gamma\gamma'\gamma''} \hat m_{\gamma''}~,
\end{equation}
which we use to rewrite Eq.~(\ref{eq:nearlythere}) as
\begin{equation}
  \hbar \hat m \times \left[ \frac{\partial}{\partial t} + ({\bf v}_s \cdot
  \nabla)\right] \hat m = \frac{\delta E}{\delta \hat m}~.
\end{equation}
Taking the cross product of the above equation with $\hat m$ and
using the fact that $\hat m$ is a unit vector we obtain
Eqs.~(\ref{eq:LLG})~and~(\ref{eq:sttcont}) for $\alpha=\beta=0$.

The term in the action in Eq.~(\ref{eq:actionwithspintorques})
that is proportional to ${\bf v}_{s}$ is physically understood as
the Berry phase acquired by the electrons as they drift through a
non-collinear magnetization texture. Such Berry phases occur
naturally in spin systems. To see this in more detail consider
first the simple case of a spin $S$ in a time-dependent unit
magnetic field $\hat m (t)$ with hamiltonian ${\mathcal H} = -
\hat m (t) \cdot \hat S$.  Suppose the magnetic field is varied
very slowly from $\hat m (t_i)$  to $\hat m (t_f)$, with $\hat m
(t_f)=\hat m(t_i)$, so that the system remains in its ground
state. One can show \cite{auerbach} that the quantum mechanical
wave function of the spin acquires a nontrivial phase factor
$e^{-i S \Omega}$. Here, $\Omega$ is the area on the unit sphere
enclosed by the path that the spin traces out as it adiabatically
follows the magnetic field. Using Stokes' theorem we write this
area as a line integral over the boundary of $\Omega$, that is,
\begin{equation}
 \Omega = \int_\Omega \hat m \cdot d\hat a \equiv \int_\Omega  \left[\nabla_{\hat  m}\times {\bf  A}(\hat m) \right]\cdot d\hat a = \int_{\partial
 \Omega}  {\bf  A}(\hat m) \cdot
 d\bm{\ell}~,
\end{equation}
with the monopole vector potential ${\bf A} (\hat m)$ determined
by Eq.~(\ref{eq:vecpotmonopole}). Note that the above also shows
that although the vector potential is only defined up to a gauge
transformation ${\bf A} \to {\bf A} - \nabla_{\hat m} \Lambda$,
with $\Lambda$ an arbitrary function of $\hat m$, the physical
quantity of interest, namely the area $\Omega$, is unaffected and
well-defined. Using the above results we observe that the term
proportional to ${\bf v}_s$ in the action in
Eq.~(\ref{eq:actionwithspintorques}) is determined by the area
that the magnetization traces out on the unit sphere.

Having found an action that reproduces the equation of motion for
the magnetization including the adiabatic spin transfer torque, we
define an energy functional
\begin{equation}
\label{eq:energyfctalwithcurrent}
  E_j [\hat m] = \int \frac{d\bx}{a^3}
 \hbar v_{s,\gamma}  A_{\gamma'}({\hat m} (\bx,t)) \nabla_\gamma m_{\gamma'}
 (\bx,t) + E[\hat m]~.
\end{equation}
Using this energy functional the equation of motion for the
magnetization direction is written as
\begin{equation}
 \frac{\partial \hat m }{\partial t} ={\hat m} \times
  \left( -\frac{\delta E_j[{\hat m}]}{\hbar \delta {\hat m}}
  \right)~.
\end{equation}
To illustrate that the energy functional in
Eq.~(\ref{eq:energyfctalwithcurrent}) is indeed the energy that is
minimized in the presence of current and as such a useful concept,
we now add dissipative terms to the above equation. In particular,
we consider the case $\alpha \neq 0$ and $\beta=0$.  In that case
we have that
\begin{equation}
 \frac{\partial \hat m }{\partial t} ={\hat m} \times
  \left( -\frac{\delta E_j[{\hat m}]}{\hbar \delta {\hat m}}
  \right) - \alpha \hat m \times \frac{\partial \hat m }{\partial
  t}~,
\end{equation}
so that, since $\alpha>0$, the energy functional $E_j[\hat m]$ is
indeed minimized according to the above equation of motion.  This
energy functional is relevant for understanding the so-called
``intrinsic pinning" of a domain wall that occurs for $\beta=0$.
``Intrinsic pinning" refers to the fact that a magnetic domain
wall in a perfect material is displaced only for a sufficiently
large applied current, and is ``pinned" otherwise.\cite{tatara}
Minimization of this energy functional yields a physically clear
explanation of intrinsic domain wall pinning.\cite{tatara,duineDW}

We end this section with some comments regarding spin valves.
Writing down an action that reproduces Slonczewski's spin-torque
directly for the spin valve case turns out to be more complicated
because of the ``discretization limit" one has to take in going
from a smooth magnetization texture to a spin valve. We speculate,
nonetheless, that the above action in principle is sufficient, as
it reproduces Eq.~(\ref{eq:sttcont}) and we have seen the
equivalence of Eq.~(\ref{eq:sttcont}) and Eq.~(\ref{eq:sttslonc}).
At this point some readers may object that it is in fact the
dissipative effective field torque proportional to $\beta$ in
Eq.~(\ref{eq:sttslonc}) which can be derived from the energy
functional
\begin{equation}
\label{eq:efffieldtorquefakeenergyfunct}
  \tilde E[{\hat m}] =-\beta g j {\hat m}_p \cdot {\hat m}~.
\end{equation}
We note, however, that this energy is odd under time reversal and
therefore not a proper energy. Only when the pinned magnet is kept
fixed as $t \mapsto -t$ it is even under time reversal . This
corresponds to regarding the pinned magnet as distinct from the
dynamical system at hand (the free magnet).

Among the many questions which remain we mention two. i) {\em Is
the energy functional from which the reactive spin-torques are
derived a practical concept?} The prefactor of the monopole vector
potential contains the spin-current evaluated in the collinear
situation. Is this useful for calculations of the Slonczewski
spin-torque efficiency?  We note that a different perspective on
this issue can be found in Ref. ~\onlinecite{stamenova}.  ii) {\em
Can useful predictive expressions be derived for the dissipative
coefficients $\alpha$ and $\beta$ of real ferromagnetic
materials?}


\section{Current-induced torques in Ferrimagnets}
\label{sec:ferrimagnets}

For systems in which spin is approximately conserved, such as
transition metals, the current-induced torque picture is
complementary to the standard bookkeeping argument of
spin-transfer, as described earlier.  For systems in which spin is
not conserved ({\em i.e.} systems with strong spin-orbit
coupling), the bookkeeping argument is no longer valid, and the
current-induced torque picture must be adopted.  Examples of such
systems include diluted magnetic semiconductors (DMS) and
rare-earth elements. The current-induced torque picture has been
previously employed for these systems, for example by Nguyen {\em
et al.} who calculate the current-driven domain wall mobility in
the DMS (Ga,Mn)As. \cite{nguyen}

Recent experiments on ferrimagnets provide an interesting test of
current-induced torque theories.  Jiang {\em et al.} consider
magnetoresistance and CIMS in a system consisting of a
ferromagnetic CoFe fixed layer and a ferrimagnetic CoGd free
layer.\cite{jiang}  In CoGd, the magnetization of the two
sublattices Co and Gd point in opposite directions.  For
temperatures above the magnetic compensation temperature $T_{MC}$,
the net magnetization points in the same direction as the Co
sublattice, while for temperatures below $T_{MC}$, it points in
the direction of the Gd sublattice. The net magnetization vanishes
at $T=T_{MC}$.

Transport is dominated by orbitals from the Co sublattice, so that
the magnetoresistance probes the relative orientation of the Co
sublattice with the ferromagnetic pinned layer.  Therefore the
magnetoresistance changes sign at $T=T_{MC}$ (so that for
$T<T_{MC}$, the magnetoresistance is negative - in this regime the
Co sublattice is antiparallel with the net magnetization).

The experimental result of key interest is that the sense of the
CIMS changes sign not at $T=T_{MC}$, but at a higher temperature.
The interpretation provided for this experimental result was that
at all temperatures, the CIMS is determined by the relative
orientation of the total angular momentum of pinned layer and free
layers.  The angular momentum $\vec L$ and magnetization $\vec M$
of a sublattice are related by ${\vec L} = {\vec M}/\gamma$, where
$\gamma=g \mu_B / \hbar$ is the gyromagnetic ratio, whose
$g$-value depends on the the degree of spin-orbit coupling. Making
the assumption that different $g$-factors can be assigned to the
Co and Gd sublattices, we can see that CoGd has the curious
property that in a certain temperature range, its magnetization
${\vec M} = {\vec M}_{\rm Co}+{\vec M}_{\rm Gd}$ and angular
momentum ${\vec L}={\vec M}_{\rm Co}/\gamma_{\rm Co}+{\vec M}_{\rm
Gd}/\gamma_{\rm Gd}$ can have opposite signs.

At temperatures $T>T_{MC}$, the magnetization and angular momentum
are parallel, and are dictated by the Co sublattice, while at very
low temperatures, the magnetization and angular momentum are also
parallel and dictated by the Gd sublattice.  In these regimes, let
us suppose that a positive current leads to parallel alignment of
the CoGd magnetization with the fixed layer magnetization.  In the
intermediate temperature range where $\vec M$ and $\vec L$ have
opposite sign, so the interpretation goes, a positive current
aligns the angular momentum $\vec L$ ({\em not} $\vec M$) with the
fixed layer magnetization.

We believe that this is an important experimental result for
understanding the basic nature of current-induced torques.  The
interpretation provided by the experimentalists extrapolates the
{\em bookkeeping theory} from one based on conservation of only
spin-angular momentum to conservation of total angular momentum.
We do not believe that this extrapolation is well motivated. For
one thing, orbital angular momentum is not even approximately
conserved in a crystal. We do not believe that the temperature at
which the sense of CIMS changes sign must equal the temperature at
which the total angular momentum changes sign.  It is true that
because of spin-orbit interactions, ferromagnets do have an
orbital angular momentum, but neither spin nor orbital angular
momentum are conserved. Indeed, understanding the magnitude of the
orbital contribution to the magnetization of metallic ferromagnets
has proved to be a challenging problem for many-body
theory.\cite{eriksson}

Our microscopic picture of current-induced torques provides some
guidance on how to think about these interesting experiments. A
systematic inclusion of orbital magnetism is provided by
current-density functional theory (CDFT).\cite{eschrig,ebert} CDFT
is an extension of DFT in which the {\em current} density (as
opposed to {\em charge} density) plays the central role, and is
able to treat many-body systems in magnetic fields of arbitrary
strength.\cite{ebert}  The exchange-correlation potential is more
complicated in CDFT than in DFT, and includes a vector potential.
However, a standard practice is to simply employ density
functional theory with the standard exchange-correlation
potentials that depend only on spin-density, even when spin-orbit
coupling is included in the Kohn-Sham single-particle equations.
One formal justification for this comes from the relativistic
spin-density-functional formalism.\cite{rsdf}  This formalism
allows for relativistic corrections to the exchange-correlation
potential, but the corrections do not have an overwhelming effect.
Taking this as a starting point, the current-induced spin-torques
in CoGd would be proportional to induced spin-densities and would
be influenced by the spin-orbit interaction terms in the Kohn-Sham
equations for the current-carrying quasiparticles.

Another aspect that appears to play a crucially important role is
the thermal fluctuations of the magnetization, something which has
not usually been accounted for by theory.  The effect of these
fluctuations is not clear {\em a priori} (and indeed represents an
interesting avenue of research in its own right). At finite
temperature, both the Gd moments and the Co moments will fluctuate
in orientation, so that the magnetization cannot be assumed to be
collinear even within the CoGd nanomagnet. Presumably the change
in sign of the average magnetization occurs for $T>T_{MC}$ because
the Gd moment orientations fluctuate more strongly.

It is interesting to consider how we would expect thermal
fluctuations in such a system to influence macroscopic
current-induced torques.  Is the explanation for these
experimental results related only to thermal fluctuations, with
spin-orbit coupling playing an inessential role?  Since the sum of
the current-induced spin-torques on individual atoms will not in
general be perpendicular to the total spin-density, its
effectiveness in driving spatially coherent precession of the
typical non-collinear spin-density is not simply related to the
net spin-current through the nanomagnet.  The intriguing
experimental results in CoGd may indeed be pointing to a
non-trivial role for thermal fluctuations of the spin-density in
current-induced torque phenomena, when these fluctuations are
large.

\section{Spintronics in Antiferromagnets}

The more general nature of the current-induced torque picture
suggests that it should be operational in more general
circumstances, for examples in materials with more complex
magnetic order than simple collinear ferromagnetism. We have
recently considered
\cite{alvaroAFM,haneyAFM,duineinelastic,haneynew,tsoiAFM} the
effect of current-induced torques in circuits containing different
combinations of antiferromagnetic (AFM) and ferromagnetic
materials. To date, the role of antiferromagnets in spintronic
devices is to pin a ferromagnetic layer's orientation via an
effect known as ``exchange bias". \cite{ex1,ex2}  We propose that
antiferromagnetic materials can serve as building block for
circuits which display effects such as giant magnetoresistance
(GMR) and CIMS, with the staggered order parameter of the AFM
playing the role of the orientation of the FM.  We briefly discuss
our results for the sake of illustrating the qualitatively new
features than can emerge from antiferromagnetic spintronics. These
early efforts may help point the way to fruitful directions to
consider in moving forward.

The property that the local spin-transfer torque on each atomic
site is equal to the net spin-current into that site is equally
true for ferromagnets and antiferromagnets.  The difference
between the two-cases is the nature of the magnetic order.  The
magnitude of current-induced torques is nearly always dwarfed by
the size of the torques caused by equilibrium exchange
interactions when the relative orientations of different moments
is distorted. In the case of a ferromagnet, equilibrium exchange
orientations keep all the moments in a nanomagnet essentially
rigidly parallel; the current-induced torques are generally not
strong enough to change these relative orientations. The sum of
all the current-induced torques acts on the overall orientation of
all the moments.  For this mode of magnetic dynamics the
current-induced torque only has to compete with much weaker
anisotropy, magnetostatic, external field, and damping effects. In
an antiferromagnet, the mode of magnetic dynamics on which the
current-induced torque can have an effect is the rigid motion of
all the moments of the antiferromagnet.  For an antiferromagnet
with two sublattices with opposite orientations, it is the
difference between the sums of the torques over the two
sublattices which has an effect.

GMR and current-induced torque (CIT) effects in ferromagnets rely
on the interplay between electron transport and magnetic order.
The source of this interplay is the strong spin-dependent
exchange-correlation potentials seen by current-carrying
quasiparticles, which is the result of a spin-dependent Fermi
surface. Antiferromagnets do not posses a spin-dependent Fermi
surface, the characteristic that is so essential to conventional
spintronics, so it is clear that any GMR and CIT effects in AFM
circuits must have a fundamentally different origin.  (Indeed the
only qualitative imprint of the ordered state on the electronic
structure of AFM is the formation of a gap at the spin-density
wave vector; we denote the spin-density wave vector by $\vec Q$ .)
We can classify our studies by the relative orientation of the
current $\vec J$ and the spin-density wave vector $\vec Q$.

\noindent {\em Case 1: ${\vec J} \parallel {\vec Q}$.} In Ref.
~\onlinecite{alvaroAFM} we considered an antiferromagnets in which
the exchange splitting changed sign on alternate lattice sites in
the current-motion direction (${\vec J} \parallel {\vec Q}$).  The
structure we studied with this kind of model was similar to that
of a normal spin valve, except that the magnetic nanoparticles
separated by a normal metal spacer were antiferromagnetic rather
than ferromagnetic.  We refer to this type of structure as an
antiferromagnetic spin valve.  The model we studied in this paper
was a single-band {\em toy model}.  In the ${\vec J}
\parallel {\vec Q}$ case the interface layer of the AFM is
uncompensated, {\em i.e.} it has nonvanishing total spin.   In
this model, we find a difference in conductivity according to
whether or not the AFM layers adjacent to the spacer are parallel
or anti-parallel, an effect we refer to as antiferromagnetic giant
magnetoresistance (AGMR). We also find the remarkable property
that when the AFM layers are noncollinear, the out-of-plane
spin-density in the AFM is exactly {\it constant} in our lattice
model antiferromagnet and exactly {\it periodic} in a continuum
model antiferromagnet. Since the out-of-plane spin-density is
responsible for the current-induced torque, the torque is also
constant in magnitude and alternating in sign throughout the
volume of the AFM. By itself, this property implies that the
critical current for switching is independent of the AFM layer
thickness. This is in stark contrast to the FM case, where the
spin-torque decays rapidly away from the interface as the result
of interference between different transverse channels'
spin-density.

Since the staggered magnetization of antiferromagnets is not
conserved, the staggered torque which drives order parameter
dynamics is not protected by robust conservation laws.  Indeed,
both the AGMR and the current-induced torques of the uncompensated
AFM spin-valve toy model can be seen to follow from
phase-coherence effects: specifically the difference in the phase
acquired by up and down spins as they traverse the circuit in
various transverse channels.  The presence of the current-induced
torque in this geometry may also be understood from the
bookkeeping perspective: the difference in phase acquired by the
up and down-component of an electron spin as it traverses a single
uncompensated layer results in a net spin flux into that layer.

The toy model calculations demonstrate that GMR can occur in
purely antiferromagnetic spin valves and that current-induced
torques can drive antiferromagnetic order parameters.  At the same
time, the toy model considerations also suggest that in the AFM
case the effects are more easily weakened by inelastic scattering
and more sensitive to the details of the various interfaces in
these layered systems.\cite{duineinelastic}

The toy model does not fully capture all the physics that is
present in realistic AFM spin-valve structures.  To partially
assess the significance of real-world complications we have
performed {\em ab initio} calculations for a system consisting of
antiferromagnetic Cr leads separated by a Au
spacer.\cite{haneyAFM}  The AGMR in this calculation is not
primarily due to phase-coherent effects, but rather to
spin-polarized interface resonance.  The [001] surface of Cr with
spin density wave in the [001] direction exhibits very different
properties than that of the bulk, with an enhanced magnetic moment
and a spin-dependent local density of states at the
surface.\cite{fu} The spin-dependence of the surface density of
states at the Fermi energy results in a spin-polarized current as
electron flow passes through the surface. The presence of Au
spacer has little effect on this surface state, and the Cr layer
can essentially be thought of similarly as a FM layer, with the
surface magnetization playing the role of the FM magnetization.

Another different aspect of this system is that the
non-equilibrium out-of-plane spin-density is not periodic as in
the simpler models, but partially decays away from the interface,
because of the complex Fermi surface of Cr. For many transverse
channels, the spin-dependent scattering state is a linear
combination of Bloch states with different wave-vectors in the
transport direction $k_z$.  The spin-densities of these scattering
states then show an oscillatory spatial structure, and averaging
over the Fermi surface results in destructive interference of the
non-equilibrium spin-density away from the interface, as in the
conventional FM case.  We address some open questions regarding
this system at the end of this section.

\begin{figure}[h]
\begin{center}
\vskip 0.25 cm
\includegraphics[width=4.in,angle=0]{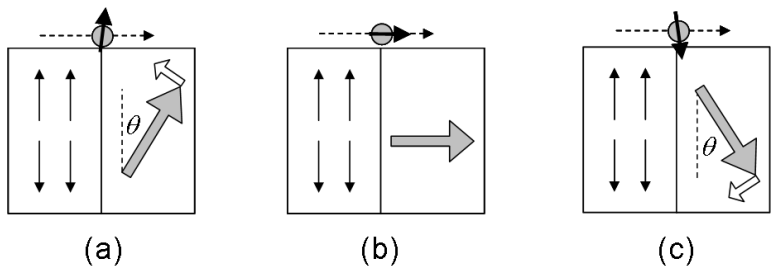}
\vskip 0.25 cm \caption{Current-induced torques due to a
compensated antiferromagnet.  The arrows above the structure
indicate the electron flux spin direction.  The white arrows
indicate the ensuing current-induced torques on the FM.  The
torque vanishes at $\theta=90^\circ$, and varies as $\sin
2\theta$.} \label{fig:AFM-FM}
\end{center}
\end{figure}

\noindent {\em Case 2: ${\vec J} \perp {\vec Q}$.} We have also
considered a system consisting of a ferromagnet and
antiferromagnet in which the current is perpendicular to the
spin-density wave ( ${\vec J} \perp {\vec Q}$).  The total
magnetization of each layer in the current-carrying direction
vanishes in this case, {\em i.e.} the magnetization is
compensated.  Most AFM materials used in magnetoelectronics are
fully compensated, {\em i.e.} the spin-density sums to zero in
every lattice plane perpendicular to the current direction, or at
least nearly so. Direct interfaces between nearly compensated
antiferromagnets, which perform the {\em exchange bias} function,
and ferromagnets are common in spintronic circuits.  The
current-induced torques discussed in this section can be dominant
when the antiferromagnetic and ferromagnetic layers are separated
by a spacer which reduces the importance of direct exchange
interactions between the two actors. Symmetry considerations imply
that the current-induced torques between ferromagnets and
compensated antiferromagnets differ qualitatively from the torques
between ferromagnets.

The total current-induced torque acting on a FM nanoparticle can
always\cite{stilesZangwill} be expressed in terms of the
difference between incoming and outgoing spin-currents.  The
presence of a ferromagnet will in general induce a nonzero
spin-current at the AFM-FM interface.  When spin-polarized
electron flux from the AFM with orientation ${\hat n}_{AFM}$
enters a FM with orientation ${\hat n}_{FM}$, the spin-current
entering the FM will have some component in the ${\hat n}_{AFM}$
direction.  It follows that, just as in the familiar case where
both materials are FMs, a current-induced torque will act in the
plane defined by ${\hat n}_{AFM}$ and ${\hat n}_{FM}$, as
illustrated in Fig. (\ref{fig:AFM-FM}). (Out of plane torques are
also non-zero but tend to be much smaller.) Spin rotational
invariance of the overall circuit implies that the in-plane torque
must be an odd function of the angle $\theta$ between
$\hat{n}_{FM}$ and $\hat{n}_{AFM}$ and that it can therefore be
expanded in terms of a $\sin$e-only Fourier series, vanishing for
both parallel and antiparallel collinear configurations.  In this
case, reversal of the AFM moment direction is equivalent to a
lateral translation which cannot influence the current-induced
torque.  It follows that in the compensated AFM case the torque is
invariant under $\theta \to \theta + \pi$, restricting its Fourier
expansion to terms proportional to $\sin(2n\theta)$.  It follows
that the torque vanishes when $\hat{n}_{FM}$ is perpendicular to
$\hat{n}_{AFM}$, and undergoes a sign change for $\theta \to \pi -
\theta$, as illustrated in Fig.(~\ref{fig:AFM-FM}).  The property
that the torque acting on a FM due to a compensated AFM vanishes
not only for collinear but also for perpendicular orientations is
primarily responsible for a novel current-induced-torque phase
diagram, which differs drastically from the now familiar
applied-field/current phase diagram for FM layers.\cite{kiselev}

We emphasize that in making this general argument of the
symmetry-constrained form of the current-induced torque on the FM,
we have appealed to the bookkeeping picture.  The bookkeeping
picture combined with the assumption that the spin current becomes
aligned to the local magnetization represents a simplification of
great conceptual and practical utility.  The calculation of the
total current-induced torque in a spin valve then requires only
the determination of the spin-current in the spacer.  On the other
hand, for the AFM it is necessary to find the staggered torque,
and this convenient picture of finding net torques from a single
spin-current value does not apply.  To date, we have found torques
on AFMs by adding up torques on individual atoms, and their form
and magnitude seemingly can not be so easily be anticipated {\em a
priori}.

We have performed a realistic calculation of the torques present
between when ferromagnetic Co is adjacent to the antiferromagnetic
compound NiMn.  As expected on the basis of the symmetry
considerations explained above, the current-induced torque is of
$\sin2\theta$ form.  We find that the current-induced torque
efficiency acting on both layers is substantial and of the same
magnitude as that found in common FM systems. For electron flow
from FM to AFM, the CIT tends to align the axis of the AFM with
that of the FM, and to make the FM perpendicular to the axis of
the AFM. Conversely, for electron flow from AFM to FM, the CIT
tends to align the FM with the AFM axis, and make the AFM axis
perpendicular to the FM (within their common plane).  Put another
way: the current-induced torque tends to drive the orientation of
the downstream material (AFM of FM) parallel with that of the
upstream, and to drive the upstream material orientation
perpendicular to the downstream.

\begin{figure}[h]
\begin{center}
\vskip 0.25 cm
\includegraphics[width=4.0in,angle=0]{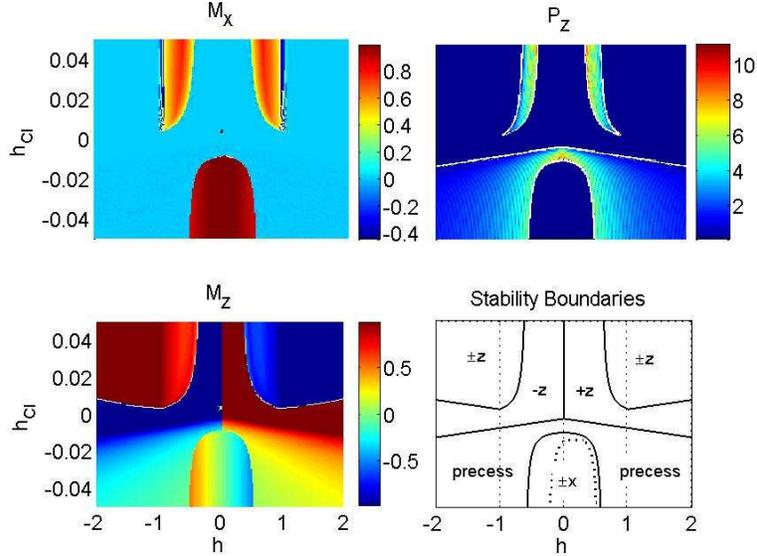}
\vskip 0.25 cm \caption{Magnetic configuration (${\rm M_x, M_z}$)
and peak of power spectrum ${\rm P_z}$ (arbitrary units) versus
applied field and current.  Also shown is stability boundaries
found analytically (the labels $\pm {\rm x}, \pm {\rm z}$ refer
also to solutions which point approximately in these directions).
The stability boundary plot also shows the reduced out-of-plane
solution space for negative to positive field sweep with a dashed
line. } \label{fig:phasePlot}
\end{center}
\end{figure}

The effect of such a torque on the stability of the FM orientation
is most unusual for electron flow from FM to AFM.  If the FM is an
easy plane ferromagnet (hard axis = $\hat x$), and the AFM axis
(AFM axis = $\hat z$) is assumed to lie in the ferromagnet's easy
plane and be fixed, then for sufficiently large current, the
stable configuration for the FM is to point approximately out of
the easy plane.  Fig. (\ref{fig:phasePlot}) shows the stable
magnetization orientation phase diagram versus applied field and
current.  The applied field is scaled by the demagnetization field
of the FM, $h=H_{app}/H_{demag}$, and the conversion of the
dimensionless current-induced torque $h_{CI}$ into a real current
$J$ (assuming a demag field of $1~T$) is $J=(h_{CI} t) \times 3.8
\cdot 10^{9} {\rm A/cm^2}$, where $t$ is the thickness of the FM
layer in nm; $h_{CI}<0$ represents particle flow from FM to AFM.

\begin{figure}[h]
\begin{center}
\vskip 0.25 cm
\includegraphics[width=3.0in,angle=0]{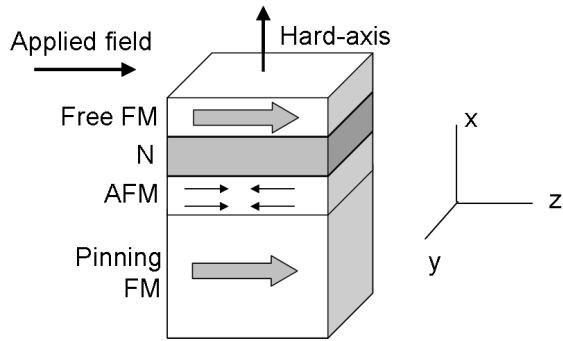}
\vskip 0.25 cm \caption{Possible stack configuration to study the
influence of compensated AFM on FM orientation as a function of
applied field and current.  The AFM is assumed to be fixed by the
larger pinning FM via the exchange bias effect. }
\label{fig:stack}
\end{center}
\end{figure}

A stack design which may be employed in future studies to
investigate this effect is shown in Fig. (\ref{fig:stack}).  The
AFM should be pinned in some way, here we've indicated pinning
with the exchange bias effect by placing an larger FM adjacent to
the AFM.  As discussed in Ref. ~\onlinecite{haneynew}, it should
not be necessary for the AFM to be single domain in order to see
the effect.  The orientation of the free FM may be detected via
the GMR effect with the pinning FM. The pinning FM may play a role
in the stability of the free FM, but its role is sufficiently
distinct from that of the AFM that it may be possible to
nevertheless detect the AFM current-induced torque.
 A detailed discussion of this unexpected behavior can be found in Ref.
~\onlinecite{haneynew}.  The qualitatively distinct nature of the
applied current-applied field phase diagram suggests that there
may be other interesting aspects of this form of torque to be
explored as well.

{\em What next for AFM spintronics?}  There are other permutations
of ${\vec J}$, ${\vec Q}$ orientation and FM/AFM stack design to
be explored.  We believe the key step necessary for making further
progress is finding materials systems which clearly demonstrates
the effect of current-induced torques on or by an AFM layer in a
way which can be distinguished from FM STT.

Experimental challenges abound for studying AFM spintronics. Chief
among them is that the structure and orientation of an AFM is
difficult to measure and difficult to control.  The easiest way to
control the AFM orientation is with an adjacent FM via the
exchange bias effect. However this FM layer may exert spin-torques
of its own, potentially obscuring the role of the AFM.  It would
be preferable therefore to avoid using a pinning FM, or to somehow
remove its effect on transport. To the extent that effects rely on
flat interfaces or phase-coherence, the experimental challenges
become more severe. Nevertheless, recent experiments\cite{tsoiAFM}
established a dependence of unidirectional exchange bias fields on
current, providing indirect evidence that current-induced torques
are present in AFMs.  Other experiments attribute observed changes
in exchange bias, steps in differential resistance, and the
statistics of thermally activated switching to current-induced
torques on the antiferromagnet of an exchange-biased spin
valve.\cite{urazhdin}

The role of disorder on both toy and more realistic models is
likely much more important in antiferromagnets than in
ferromagnets.  This role has not yet been realistically assessed,
and represents an important step forward in determining the
viability of AFM spintronics which can be achieved by theoretical
work alone. For example, some important properties of the ${\vec
J} \parallel {\vec Q}$ geometry seem to rely to some extent on a
clean, uncompensated AFM surface.  The robustness of the GMR and
CIT when this theoretical assumption is relaxed is still largely
unknown.

Finally, so far we've only considered the effect of current on
AFMs in multilayer geometries.  It remains to study the effect of
current on continuous AFM textures and domain walls, although some
work is currently underway.\cite{xia} Domain walls in AFMs display
interesting properties in their own right, including evidence of
quantum tunnelling.\cite{shpyrko} Clearly there is a wide range of
phenomena in AFM spintronics that is yet to be explored.

\section{Spintronics in molecular systems/Pseudospintronics}

In the previous two sections, new and interesting physics was
revealed by considering CIT effects in materials other than FM
(ferrimagnets and antiferromagnets).  That provokes the question
of what {\em other} types of materials or systems might provide
interesting manifestations of CIT.

One type of system in which CIT effects may occur in interesting
ways is in molecular systems, or those with reduced
dimensionality.  So far experiments have focused on
magnetoresistance effects for systems with ferromagnetic leads
sandwiching monolayers of molecules \cite{petta} or carbon
nanotubes,\cite{tsukagoshi} for example. The non-equilibrium
Green's function is well suited to calculating spin-dependent
transport for such systems.\cite{waldron}  A next step would be to
consider the action of current on magnetic molecules to see if
current-induced switching of the molecular spin is possible, and
to consider experimental signatures of such a switching
event.\cite{misiorny,fransson} Experimental challenges are
abundant, and certainly outside of the expertise of these authors,
however first-principles calculations may be useful in aiding the
identification of the optimum experimental choice for molecule and
lead material.  There can however be difficulties in applying
density functional theory (strictly speaking a {\em ground state}
mean-field theory) directly to molecular transport (a non-ground
state system in which electron correlations may be
important).\cite{nitzan,evers,sai}  It is nevertheless likely that
the current-induced torques in such systems will be qualitatively
different than in conventional FMs. In FMs, the properties of
current-induced torques rely critically on the dimensionality (the
spin-current becomes aligned to the local magnetization only after
averaging over all incoming electron velocity directions), and
these molecular systems are effectively 1-d or 0-d.

Finally we mention that current-induced torques are closely
related to physics that occurs in other kinds of systems.  For
example, the supercurrent that flows through a superconductor in a
circuit with a normal metal source and drain can be thought of as
being driven by current-induced changes in the equation of motion
for the order parameter.  In this case, total particle number (a
scalar) rather than total spin (a vector) is conserved.  The case
of circuits containing superconductors is therefore closely
analogous to the case of XY easy-plane ferromagnets for which only
the $\hat{z}$-spin is conserved.  Bilayer quantum Hall systems
near filling factor $\nu=1$, have an exciton
condensate\cite{jpeahmnature} ground state with spontaneous phase
coherence in the two layers.  The conserved quantity in this case
is the difference between the numbers of particles in the two
layers. The anomalous transport properties of these exciton
superfluids are closely related\cite{rossi} to the anomalous
transport properties of magnetic systems that we have discussed in
this review.  We suspect that many other examples of
current-driven order parameters will be discovered and exploited
in the future.
So far, metallic magnetism has provided the most
phenomenologically rich and most extensively explored example of
this type of physics.  The lessons learned from this still
developing body of research may have implications beyond the realm
of magnetism.

\section{Conclusion}

In this review, we have described a microscopic theory of
current-induced torques which identifies the interaction between
misaligned spins of non-equilibrium quasiparticles and the
magnetic condensate as the source of the torque.  This perspective
suggests that the phenomena of current-driven order parameters is
more general than the spin-transfer idea.  It suggests that in
magnetic systems the phenomena is not limited to ferromagnets, or
to systems in which total spin is conserved.  We have applied this
picture to consider spintronics in antiferromagnets, and have
found a number of qualitative differences in the physics of CITs
compared to the ferromagnetic case.  There are a number of new
areas that have not yet been fully explored with this approach -
among them systems with strong spin-orbit coupling, different
types of antiferromagnetic systems, and molecular systems.  It is
our hope that in exploring these novel systems, the key,
fundamental aspects of current-induced torques can be illuminated
in their most general form.

We would like to acknowledge stimulating discussions with Hong
Guo, Olle Heinonen, Enrico Rossi, Neal Smith, Maxim Tsoi, and
Derek Waldron.

The work of R.A.D. is partially supported by the Stichting voor
Fundamenteel Onderzoek der Materie (FOM) and the Nederlandse
Organisatie voor Wetenschappelijk Onderzoek (NWO).  The work of
A.S.N. is partially supported by Proyecto Bicentenario de Ciencia
y Tecnolog\'ia ACT027.  Work at UT Austin was supported in part by
Seagate Corporation and by the National Science Foundation under
grant DMR-0606489.

\end{document}